\documentclass{appolb}
\usepackage{graphicx}
\usepackage{times}
\usepackage[english]{babel}
\usepackage{amsmath}
\usepackage[latin1]{inputenc}
\usepackage[T1]{fontenc}
\usepackage{graphicx}
\usepackage{epsfig}
\usepackage{rotating}
\usepackage{subfigure}
\usepackage{tikz}
\usepackage[subfigure]{ccaption}
\usepackage{mathtools,cancel}

\newcommand{\dt}{{\mathrm{det}}}

\newcommand{\ud}{\mathrm{d}}
\newcommand{\st}{{\mathrm{st}}}

\def\ii{\'{\i}}

\begin{document}
\title{The 3 f\mbox{}lavor NJL with explicit symmetry breaking interactions:
scalar and pseudoscalar spectra and decays.\thanks{Presented at the workshop "Eef70" on Unquenched Hadron Spectoscopy: Non-Perturbative Models and Methods of QCD vs. Experiment, 
01-05 September 2014 in Coimbra, Portugal. Work supported by Centro de F\ii sica Computacional (CFC) da Universidade de Coimbra.
Part of the EU Research Infrastructure Integrating Activity 
Study of Strongly Interacting Matter (HadronPhysics3) under the 7th Framework 
Programme of EU: Grant Agreement No. 283286.}%
}
\author{A. A. Osipov, B. Hiller, A. H. Blin
\address{Departamento de F\ii sica, CFC, Faculdade de Ci\^encias e Tecnologia da Universidade de Coimbra, P-3140-308 Coimbra, Portugal}
}
\maketitle
\begin{abstract}

The effective quark interactions that break explicitly the chiral $SU(3)_L\times SU(3)_R$ and $U_A(1)$ symmetries by current-quark mass source terms are considered in NLO in $N_c$ counting. They are of the same order as the 't Hooft flavor determinant and the eight quark interactions that extend the LO Nambu-Jona-Lasinio Lagrangian, and complete the set of non-derivative and spin 0 interactions relevant for the $N_c$ scheme. 
The bosonized Lagrangian at meson tree level describes accurately the empirical ordering and magnitude of the splitting of states in the low lying pseudoscalar and scalar meson nonets, for which the explicit symmetry breaking terms turn out to be essential. 
The strong interaction and radiative decays of the scalar mesons are understood in terms of the underlying  microscopic multi-quark states, which are probed differently by the strong and the electromagnetic interactions. We also obtain that the anomalous two photon decays of the pseudoscalars are in very good agreement with data.
\end{abstract}

\vspace{0.5cm}

Effective low energy Lagrangians of QCD are operational at the scale of spontaneous breaking of chiral symmetry, of the order of $\Lambda_{\chi SB}\sim 4\pi f_{\pi}$ \cite{Georgi:1984}. In the Nambu-Jona-Lasinio (NJL) model \cite{Nambu:1961} this scale is also related to the gap equation and given by the ultra-violet cutoff $\Lambda$ of the one-loop quark integral, above which one expects non-perturbative effects to be of less importance. 
We consider in our Lagrangian \cite{Osipov:2013a,Osipov:2013b} generic vertices $L_i$ of non-derivative type that contribute to the effective potential as $\Lambda\to\infty$\begin{equation}
\label{genL}
   L_i\sim \frac{\bar g_i}{\Lambda^\gamma}\chi^\alpha\Sigma^\beta,
\end{equation} 
where powers of $\Lambda$ give the correct dimensionality of the interactions (below we use also unbarred couplings, $g_i=\frac{\bar g_i}{\Lambda^\gamma}$); the $L_i$ are C, P, T and chiral $SU(3)_L\times SU(3)_R$ invariant blocks, built of powers of the  sources $\chi$
which at the end give origin to the explicit symmetry breaking and have the same transformation properties as the $U(3)$Lie-algebra valued field $\Sigma =(s_a-ip_a)
\frac{1}{2}\lambda_a$; here $s_a=\bar q\lambda_aq$, $p_a=\bar q\lambda_ai
\gamma_5q$, and $a=0,1,\ldots ,8$, $\lambda_0=\sqrt{2/3}\times 1$, $\lambda_a$ 
being the standard $SU(3)$ Gell-Mann matrices for $1\leq a \leq 8$.

The interaction Lagrangian without external sources $\chi$ is well known, 
\begin{eqnarray}
\label{L-int}
   L_{int}&=&\frac{\bar G}{\Lambda^2}\mbox{tr}\left(\Sigma^\dagger\Sigma\right)
   +\frac{\bar\kappa}{\Lambda^5}\left(\det\Sigma+\det\Sigma^\dagger\right) 
   \nonumber \\
   &+&\frac{\bar g_1}{\Lambda^8}\left(\mbox{tr}\,\Sigma^\dagger\Sigma\right)^2
   +\frac{\bar g_2}{\Lambda^8}\mbox{tr}
   \left(\Sigma^\dagger\Sigma\Sigma^\dagger\Sigma\right).
\end{eqnarray}  
The second term is the 't Hooft determinant \cite{Hooft:1976}, {\cite{Bernard:1988}-\cite{Dmitrasinovic:2001}, the last two the 8 quark ($q$) interactions \cite{Osipov:2006b} which complete the number of relevant vertices in 4D for dynamical chiral symmetry breaking \cite{Andrianov:1993a}.  
The interactions dependent on the sources $\chi$  contain eleven terms \cite{Osipov:2013a,Osipov:2013b},
\begin{equation}
   L_\chi =\sum_{i=0}^{10}L_i,
\end{equation}
\begin{eqnarray}
\label{L-chi-1}
   L_0&=&-\mbox{tr}\left(\Sigma^\dagger\chi +\chi^\dagger\Sigma\right), \hspace{0.5cm}
   L_1=-\frac{\bar\kappa_1}{\Lambda}e_{ijk}e_{mnl}
   \Sigma_{im}\chi_{jn}\chi_{kl}+h.c.
   \nonumber \\
   L_2&=&\frac{\bar\kappa_2}{\Lambda^3}e_{ijk}e_{mnl}
   \chi_{im}\Sigma_{jn}\Sigma_{kl}+h.c., \hspace{0.5cm}
   L_3=\frac{\bar g_3}{\Lambda^6}\mbox{tr}
   \left(\Sigma^\dagger\Sigma\Sigma^\dagger\chi\right)+h.c.
   \nonumber \\
   L_4&=&\frac{\bar g_4}{\Lambda^6}\mbox{tr}\left(\Sigma^\dagger\Sigma\right)
   \mbox{tr}\left(\Sigma^\dagger\chi\right)+h.c., \hspace{0.5cm}
   L_5=\frac{\bar g_5}{\Lambda^4}\mbox{tr}\left(\Sigma^\dagger\chi
   \Sigma^\dagger\chi\right)+h.c.
   \nonumber \\
   L_6&=&\frac{\bar g_6}{\Lambda^4}\mbox{tr}\left(\Sigma\Sigma^\dagger\chi
   \chi^\dagger +\Sigma^\dagger\Sigma\chi^\dagger\chi\right), \hspace{0.5cm}
   L_7=\frac{\bar g_7}{\Lambda^4}\left(\mbox{tr}\Sigma^\dagger\chi 
   + h.c.\right)^2
   \nonumber \\ 
   L_8&=&\frac{\bar g_8}{\Lambda^4}\left(\mbox{tr}\Sigma^\dagger\chi 
   - h.c.\right)^2, \hspace{0.5cm}
   L_9=-\frac{\bar g_9}{\Lambda^2}\mbox{tr}\left(\Sigma^\dagger\chi
   \chi^\dagger\chi\right)+h.c.
   \nonumber \\
   L_{10}&=&-\frac{\bar g_{10}}{\Lambda^2}\mbox{tr}\left(\chi^\dagger\chi\right)
   \mbox{tr}\left(\chi^\dagger\Sigma\right)+h.c.
\end{eqnarray}
The $N_c$ assignments are
$\Sigma \sim N_c$; $\Lambda \sim N_c^0 \sim 1$; $\chi \sim N_c^0 \sim 1$ \footnote{The counting for $\Lambda$ is a direct consequence of the gap equation $1\sim N_c G\Lambda^2$.}.
We get that exactly the diagrams which survive as $\Lambda\rightarrow\infty$ also surive as $N_c\rightarrow\infty$ and comply with the usual requirements.

At LO in $1/N_c$ only the  $4q$ interactions $(\sim G)$ in eq. (\ref{L-int}) and $L_0$ contribute.
The Zweig's rule violating vertices are always of order $\frac{1}{N_c}$ with respect to the leading contribution. Non OZI-violating Lagrangian pieces scaling as $N_c^0$ represent NLO contributions with one internal quark loop in 
$N_c$ counting; their couplings encode the admixture of a four quark component ${\bar q}q{\bar q}q$ to the leading ${\bar q}q$ at $N_c\rightarrow\infty$.
Diagrams tracing Zweig's rule violation are: $\kappa,\kappa_1,\kappa_2,g_1,g_4,g_7,g_8,g_{10}$;
Diagrams with admixture of 4 quark and 2 quark states are: $g_2,g_3,g_5,g_6,g_9$.

With all the building blocks in conformity with the symmetry content of 
the model, one is free to choose the external source $\chi$. Putting $\chi 
=\frac{1}{2}\mbox{diag}(\mu_u, \mu_d, \mu_s)$,  
we obtain a consistent set of explicitly breaking chiral symmetry terms. 

From the 18 model parameters, 3 of them ($\bar\kappa_1, 
\bar g_9, \bar g_{10}$) contribute to the current quark masses $m_i, i=u,d,s$ and express the Kaplan-Manohar ambiguity \cite{Manohar:1986}. They can be set to 0 without loss of generality. One ends up with 5 parameters needed to describe the LO contributions (the scale $\Lambda$, the coupling $G$, and the $m_i$)  and 10 in NLO ( $\bar\kappa, \bar
\kappa_2$, $\bar g_1,\ldots,\bar g_{8}$). 
They are controlled on the theoretical side through the symmetries of the Lagrangian and on the phenomenological side through the low energy characteristics of the pseudoscalar and the scalar mesons.

The details of bosonization in the framework of functional integrals, which lead finally from  $L=\bar q i\gamma^\mu\partial_\mu q+L_{int}+L_{\chi}$
to the long distance ef\mbox{}fective 
mesonic Lagrangian ${\cal L}_{{\rm bos}}$, can be found in \cite{Osipov:2001},\cite{Osipov:2004b},\cite{Osipov:2013a,Osipov:2013b},

\begin{eqnarray}
\label{bos}
    &&{\cal L}_{{\rm bos}}={\cal L}_{\st}+{\cal L}_{{\rm hk}}, \nonumber \\
    &&{\cal L}_{\st}=h_a\sigma_a+\frac{h_{ab}^{(1)}}{2} 
                  \sigma_a\sigma_b+\frac{h_{ab}^{(2)}}{2} 
                  \phi_a\phi_b +\sigma_a(\frac{1}{3}+h^{(1)}_{abc}\sigma_b\sigma_c
   +h^{(2)}_{abc}\phi_b\phi_c)+ \ldots  
                  \nonumber\\
    &&W_{{\rm hk}}(\sigma,\phi )=\frac{1}{2}\mbox{ln}|\dt 
                 D^{\dagger}_E D_E|=-\!\int\!
                 \frac{\ud^4 x_E}{32 \pi^2}
                 \sum_{i=0}^\infty I_{i-1}\mbox{tr}(b_i)=\int\!\ud^4 x_E
                 {\cal L}_{{\rm hk}},
                 \nonumber\\
    &&b_0 =1,\hspace{0.2cm} b_1=-Y,
                 \hspace{0.2cm} b_2=\frac{Y^2}{2} 
                 +\frac{\lambda_3 }{2}\Delta_{ud}
                 Y
                 +\frac{\lambda_8}{2\sqrt{3}} (\Delta_{us}+\Delta_{ds})
                 Y, \hspace{0.2cm} \ldots , 
                 \nonumber\\
    &&Y=i\gamma_{\alpha}(\partial_{\alpha}\sigma 
                 +i\gamma_5\partial_{\alpha}\phi )
                 +\sigma^2+\{{\cal M},\sigma\}+\phi^2 
                 +i\gamma_5[\sigma+{\cal M},\phi ]
\end{eqnarray}
with $\Delta_{ij}=M_i^2-M_j^2$. Here $\sigma=\lambda_a\sigma_a$ and 
$\phi=\lambda_a\phi_a$ are nonet valued scalar and pseudoscalar f\mbox{}ields. The ${\cal L}_\st$ is the result of the 
stationary phase integration at leading order, over the auxiliary bosonic variables $s_a,p_a$, shown in (\ref{bos}) as
a series in growing powers of $\sigma_a$ and $\phi_a$. The coefficients $h_{ab...}$ in ${\cal L}_{\st}$ are obtained recursively from $h_a$ (which are related to the condensates).   
The result of the remaining 
Gaussian integration over the quark f\mbox{}ields is given by $W_{\rm hk}$, in the heat kernel approach. 
The Laplacian in euclidean space-time ${D^{\dagger}_E D}_E={\cal M}^2-
\partial_\alpha^2+Y$ is associated with the euclidean Dirac operator 
$D_E=i\gamma_\alpha\partial_\alpha -{\cal M}-\sigma -i\gamma_5\phi$.
The constituent quark mass matrix is denoted by ${\cal M}=\mbox{diag}(M_u,M_d,M_s)$ 
(f\mbox{}ields $\sigma_a, \phi_a$ have vanishing 
vacuum expectation values in the spontaneously broken phase).  
The quantities $I_i$  are the arithmetic averages $I_i=\frac{1}{3}\sum_{f=u,d,s}J_i(M_f^2)$
over the 1-loop euclidean momentum integrals $J_i$ with $i+1$ vertices 
($i=0,1,\ldots$) 
\begin{equation} 
\label{Ji}
    J_i(M^2)=16\pi^2\Gamma (i+1)\!\int 
             \frac{\ud^4p_E}{(2\pi)^4}\,\hat\rho_\Lambda 
             \frac{1}{(p_E^2+M^2)^{i+1}}, 
\end{equation}  
evaluated with a Pauli--Villars 
regulator $\hat\rho_\Lambda$ with two subtractions in the integrand. Note that the integrals $I_i$ do not depend on external momenta, and thus are free from $q\bar q$ thresholds \cite{Blin:1990}. The possible external momentum dependence of an amplitude is converted to terms involving derivative interactions in ${\cal L}_{{\rm hk}}$. 
We consider only the dominant contributions to the heat kernel series, up to $b_2$ for the meson spectra and strong decays. These involve the quadratic and logarithmic in $\Lambda$ quark loop integrals $I_0$ and $I_1$ respectively. We stress that all symmetries are respected in the process of truncation, as the heat kernel series remains an invariant order by order.

\begin{table*}
\caption{\small The pseudoscalar and scalar mass spectra, the weak decay constans (all in MeV) and the mixing angles $\theta_P=-12^{\circ *}$ and $\theta_S=27.5^{\circ *}$.}
\label{table-1}
\begin{tabular*}{\textwidth}{@{\extracolsep{\fill}}lrrrrrrrrrl@{}}
\hline
     & \multicolumn{1}{c}{$m_\pi$}
     & \multicolumn{1}{c}{$m_K$}
     & \multicolumn{1}{c}{$m_\eta$}
     & \multicolumn{1}{c}{$m_{\eta'}$}
     & \multicolumn{1}{c}{$f_\pi$}
     & \multicolumn{1}{c}{$f_K$}
     & \multicolumn{1}{c}{$m_\sigma$}
     & \multicolumn{1}{c}{$m_\kappa$}
     & \multicolumn{1}{c}{$m_{a_0}$}
     & \multicolumn{1}{c}{$m_{f_0}$} \\
\hline
     & 138* & 494* & 547* & 958*  & 92*  & 113* &550 & 850* &980* & 980*  \\
\hline
\end{tabular*}
\end{table*}

\begin{table*}
\caption{\small The model parameters \small $\hat m=m_u=m_d, m_s$, and $\Lambda$ are given in 
         MeV. The couplings have the following units: $[G]=$ GeV$^{-2}$, 
         $[\kappa ]=$ GeV$^{-5}$, $[g_1]=[g_2]=$ GeV$^{-8}$. We also show here 
         the values of constituent quark masses $\hat M$ and $M_s$ in MeV.}
\label{table-2}
\begin{tabular*}{\textwidth}{@{\extracolsep{\fill}}lrrrrrrrrrrrrrrl@{}}
\hline
 & \multicolumn{1}{c}{$\hat m$} 
     & \multicolumn{1}{c}{$m_s$}
     & \multicolumn{1}{c}{$\hat M$}                         
     & \multicolumn{1}{c}{$M_s$}                             
     & \multicolumn{1}{c}{$\Lambda$}    
     & \multicolumn{1}{c}{$G$}  
     & \multicolumn{1}{c}{$-\kappa$} 
     & \multicolumn{1}{c}{$g_1$}    
     & \multicolumn{1}{c}{$g_2$}  \\ 
\hline
  & 4.0* & 100* & 373 & 544 & 828  & 10.48  & 122.0   & 3284 &173* \\  
\hline
\end{tabular*} 
\end{table*}  

\begin{table*}
\caption{\small Explicit symmetry breaking interaction couplings. The couplings have
the following units: $[\kappa_1]=$ GeV$^{-1}$, $[\kappa_2]=$ GeV$^{-3}$, 
$[g_3]=[g_4]=$ GeV$^{-6}$, $[g_5]=[g_6]=[g_7]=[g_8]=$ GeV$^{-4}$,
$[g_9]=[g_{10}]=$ GeV$^{-2}$.}
\label{table-3}
\begin{tabular*}{\textwidth}{@{\extracolsep{\fill}}lrrrrrrl@{}}
\hline
      & \multicolumn{1}{c}{$\kappa_2$}    
      & \multicolumn{1}{c}{$-g_3$}  
      & \multicolumn{1}{c}{$-g_4$} 
      & \multicolumn{1}{c}{$g_5$} 
      & \multicolumn{1}{c}{$-g_6$}   
      & \multicolumn{1}{c}{$-g_7$} 
      & \multicolumn{1}{c}{$g_8$}  \\ 
\hline
 & 6.17  & 6497 & 1235 & 213 & 1642  & 13.3  & -64   \\   
\hline
\end{tabular*}
\end{table*}   
 
\begin{table*}
\caption{\small Strong decays of the scalar mesons, $m_R$ is the resonance mass in
MeV, $\Gamma^{BW}$ and $\Gamma^{Fl}$ are the Breit-Wigner width and the Flatt\'e
distribution width in MeV, $R^S=\frac{{\bar g}^S_K}{{\bar g}_\beta}$. The couplings ${\bar g}_\beta,{\bar g}^S_K$ are dimensionless and correspond to the shown transitions $S\rightarrow PP$   and to $S\rightarrow {\bar K}K$ respectively \cite{Osipov:2013b}.   
 }
\label{table-4}
\begin{tabular*}{\textwidth}{@{\extracolsep{\fill}}lrrrrrrrl@{}}
\hline
      & \multicolumn{1}{c}{Decays}
      & \multicolumn{1}{c}{$m_R$}
      & \multicolumn{1}{c}{$\Gamma^{BW}$}
      & \multicolumn{1}{c}{$\Gamma^{Fl}$}
      & \multicolumn{1}{c}{${\bar g}_\beta$}
      & \multicolumn{1}{c}{${\bar g}^S_K$}
      & \multicolumn{1}{c}{$R^S$}
\\
\hline
      &$\sigma\to\pi\pi$ &550  &461 &    & 1.94 &0.63 & 0.33   \\
      &$f_0\to\pi\pi$    &980  &62  & 30 & 0.23 &0.30 & 3.90   \\
      &$\kappa\to K\pi$  &850  &310 &    & 1.2  & 0   &        \\
      &$a_0\to\eta\pi$   &980  &420 &46  & 1.32 &2.73 & 2.07   \\
\hline
\end{tabular*}
\end{table*}



\begin{table*}
\caption{\small Anomalous decays $\Gamma_{P\gamma\gamma}$ in KeV,
corresponding to $\theta_P=-12^\circ$, $m_R$ is the particle mass in MeV. }
\label{table-6}
\begin{tabular*}{\textwidth}{@{\extracolsep{\fill}}lrrrl@{}}
\hline
Decays    & \multicolumn{1}{c}{$m_R$}
          & \multicolumn{1}{c}{$\Gamma_{P\gamma\gamma}$}
          & \multicolumn{1}{c}{$\Gamma^{exp}_{P\gamma\gamma}$ \cite{PDG}} \\
\hline
 $\pi^0\to\gamma\gamma$  &136  &0.00798 & $0.00774637\div 0.00810933$ \\
 $\eta\to\gamma\gamma$   &547  & 0.5239 & $(39.31\pm 0.2)\%\,\Gamma_{\mbox{tot}}
                                          =0.508\div 0.569 $ \\
 $\eta'\to\gamma\gamma$  &958  &5.225   & $(2.18\pm 0.08)\%\,\Gamma_{\mbox{tot}}
                                          =3.99\div 4.70$ \\
\hline
\end{tabular*}
\end{table*}
In the following we consider the isospin limit ${\hat m}=m_u=m_d\ne m_s$.
The low lying characteristica of the spin $0$ mesons in Table \ref{table-1} and $m_i$ in Table \ref{table-2} are used as input (marked by *) to obtain the parameters indicated in Tables \ref{table-2},\ref{table-3} (for other sets, related to slightly different values of $m_\sigma(500)$, $\theta_P$ and $\theta_S$ see \cite{Osipov:2013b}).  The calculated 
values of quark condensates are: $-\langle\bar uu
\rangle^{\frac{1}{3}}=232$\ MeV, and $-\langle\bar ss\rangle^{\frac{1}{3}}=206$\ 
MeV.  We stress that without the new explicit symmetry breaking terms the high accuracy achieved for the observables had not been possible. We find that the couplings $g_8$ and $\kappa_2$ are crucial for the high precision within the pseudoscalar sector.  
Furthermore the low lying scalar nonet mesons can be obtained according to the empirical ordering: $m_\kappa<m_{a_0}\simeq m_{f_0}$,  in contrast to the $m_\sigma<m_{a_0}<m_\kappa <m_{f_0}$ sequence obtained otherwise in the framework of the NJL models, e.g. \cite{Weise:1990,Osipov:2006b,Osipov:2004b,Volkov:1984,Su:2007}.
The main parameter responsible for the lower mass of $\kappa(800)$ as compared to the mass of $a_0(980)$ is $g_3$; $g_6$ allows for fine tuning. We understand the empirical masses inside the light scalar 
nonet as a consequence of some predominance of the explicit chiral symmetry breaking terms over
the dynamical chiral symmetry breaking ones for certain states. Note that the couplings $g_3$ and $g_6$ encode ${\bar q}q{\bar q}q$ admixtures to the ${\bar q}q$ states. This establishes a link between the asymptotic meson states obtained from the effective multiquark interactions considered to the successful approaches which support $\bar qq$ states 
with a meson-meson admixture \cite{Beveren:1986} or mixing of $q\bar q$-states with 
$q^2\bar q^2$  \cite{Jaffe:1977}.

In Table \ref{table-4} are shown the strong decay widths of the scalars, which are within the current
expectations. The widths of the $a_0(980)\to\pi\eta$ and $f_0(980)\to\pi\pi$ decays are well
accomodated within a Flatt\'e description. We corroborate other model calculations
in which the coupling to the $K \bar K$ channel is needed for the description of these decays.
We obtain however that although the $a_0(980)$ meson couples with a large strength of the
multi-quark components to the two kaon channel in its strong decay to two
pions, it evidences a dominant $q\bar q$ component in its radiative decay. The latter is thus fairly well described by a quark 1-loop triangle diagram, $\Gamma_{a_0\gamma\gamma}= 0.38$ KeV.
As opposed to this, the $\sigma$ and $f_0(980)$ mesons do not display an enhanced
$q\bar q$ component neither in their two photon decays nor strong decays. The quark 1-loop contributions $\Gamma_{f_0\gamma\gamma}= 0.08$ KeV,  
$\Gamma_{\sigma\gamma\gamma}= 0.21$ KeV, fall short of describing the data.  
Finally, the anomalous 2 photon decays of the pseudoscalars are in very good
agreement with data, see Table \ref{table-6}. For a full discussion see \cite{Osipov:2013b}.  

The response to the external parameters $T,\mu$ has been recently addressed in \cite{Moreira:2014}, with implications on strange quark matter formation.


\end{document}